\documentclass{ptapap}

\newcommand{\rhoBH}{\ensuremath{\rho_\mathrm{BH}}}
\newcommand{\rhoBHunits}{\ensuremath{\,\times 10^5\Msun\MPC\3}}

\newcommand{\eff}{\ensuremath{\epsilon}}

\newcommand{\Msun}{\ensuremath{~\mathrm{M}_\odot}}

\newcommand{\Mdot}{\ensuremath{\dot M}}

\newcommand{\zo}{\ensuremath{z_s}}

\newcommand{\3}{\ensuremath{^{-3}}}

\newcommand{\MPC}{\ensuremath{\mathrm{\,Mpc}}}

\def\simlt{\mathrel{\rlap{\lower 3pt\hbox{$\sim$}}\raise 2.0pt\hbox{$<$}}}
\def\simgt{\mathrel{\rlap{\lower 3pt\hbox{$\sim$}} \raise
2.0pt\hbox{$>$}}}
\def\lsim{\,\lower2truept\hbox{${<\atop\hbox{\raise4truept\hbox{$\sim$}}}$}\,}
\def\gsim{\,\lower2truept\hbox{${> \atop\hbox{\raise4truept\hbox{$\sim$}}}$}\,}

\author{Gianfranco De Zotti}[OaPd]
\affil[OaPd]{INAF-Osservatorio Astronomico di Padova, \\ Vicolo dell'Osservatorio 5, I-35122 Padova, Italy
}

\title{Co-evolution of galaxies and Active Galactic Nuclei}

\begin{document}

\maketitle

\def\simlt{\mathrel{\rlap{\lower 3pt\hbox{$\sim$}}\raise 2.0pt\hbox{$<$}}}
\def\simgt{\mathrel{\rlap{\lower 3pt\hbox{$\sim$}} \raise
2.0pt\hbox{$>$}}}
\def\lsim{\,\lower2truept\hbox{${<\atop\hbox{\raise4truept\hbox{$\sim$}}}$}\,}
\def\gsim{\,\lower2truept\hbox{${> \atop\hbox{\raise4truept\hbox{$\sim$}}}$}\,}

\begin{abstract}
Supermassive black holes (SMBHs) have been found to be ubiquitous in the
nuclei of early-type galaxies and of bulges of spirals. There are evidences
of a tight correlation between the SMBH masses, the velocity dispersions of
stars in the spheroidal components galaxies and other galaxy properties. Also
the evolution of the luminosity density due to nuclear activity is similar to
that due to star formation. All that suggests an evolutionary connection
between Active Galactic Nuclei (AGNs) and their host galaxies. After a review
of these evidences this lecture discusses how AGNs can affect the host
galaxies. Other feedback processes advocated to account for the differences
between the halo and the stellar mass functions are also briefly introduced.

\end{abstract}

\section{Introduction}

The mutual interactions between super-massive black holes (SMBHs) and host
galaxies are a key ingredient to understand the evolution of both source
populations. There are now clear evidences that SMBHs and host galaxies evolve
in a coordinated way over cosmic history. The understanding of the processes
that drive the SMBH-galaxy co-evolution is a central topic in current
extragalactic studies.

The starting point was the discovery, via stellar/gas dynamics and photometric
observations, that nearby galaxies possessing massive spheroidal components
(bulges) host, at their centers, a massive dark object (MDO) endowed with a
mass proportional to the mass in old stars or to the K-band luminosity
\citep{KormendyRichstone1995, Magorrian1998}. The MDOs were generally
interpreted as the black hole (BH) remnants of a past nuclear activity.
\citet{Salucci1999} demonstrated that the mass function of MDOs, derived on the
basis of the observed correlation between their masses and the stellar masses
in galactic bulges, is consistent with the SMBH mass function derived from the
redshift-dependent luminosity function of Active Galactic Nuclei (AGNs), for
standard values of the radiative efficiency of BH accretion.

The SMBH interpretation of MDOs was strongly confirmed by the analysis of the
orbits for a number of individual stars in the central region of our Galaxy
\citep{Ghez2008, Genzel2010}, ruling out alternative possibilities. A further
quite unambiguous evidence of a central SMBH was provided by VLBI measurements
at milli-arcsecond (mas)  resolution of the $H_2 O$ mega-maser at 22\,GHz of
the galaxy NGC\,4258 \citep{Miyoshi1995}.

The luminosity/mass of the stellar component is not the only global property of
the local ETGs that correlates with the central BH mass, $M_{\rm BH}$. In fact,
a tighter correlation was found between $M_{\rm BH}$  and the stellar velocity
dispersion \citep{FerrareseMerritt2000, Gebhardt2000}. The connection between
SMBH mass and galaxy properties, linking scales differing by as much as nine
orders of magnitude is likely imprinted by the huge amount of energy released
by AGNs, a small fraction of which may come out in a mechanical form (AGN
feedback). However, many important details of the processes governing the
AGN-galaxy co-evolution are still to be clarified.

In this paper, after a short introduction to AGNs (Sect.~\ref{sect:AGNintro}) I
will touch on their demography (Sect.~\ref{sect:BHdemography}).
Section~\ref{sect:Soltan} deals the main mechanism of SMBH growth: radiative
accretion and/or non-radiative processes such as BH mergers? The argument put
forward long ago by \citet{Soltan1982} has still an important bearing on this
matter. Section~\ref{sect:interaction} is about relationships between galaxy
properties and the central SMBH. Section~\ref{sect:coev} concerns the AGN
impact on the host galaxy; the two main mechanisms corrently discussed in the
literature, energy- and momentum-driven winds, are illustrated. In
Section~\ref{sect:feedback} winds powered by AGNs are put in the more general
context of feedback processes invoked to account for the different shapes of
the mass functions of dark matter halos and of galaxies. Finally, the main
conclusions are summarized in Sect.~\ref{sect:conclusions}.

\section{Introduction to AGNs}\label{sect:AGNintro}

The AGN (or quasar) discovery dates back to the \citet{Schmidt1963} paper in
which the redshift measurement of a bright radio source, 3C\,273, was reported:
``The stellar object is the nuclear region of a galaxy with a cosmological
redshift of 0.158, corresponding to an apparent velocity of 47,400 km/s. The
distance would be around 500 megaparsecs, and the diameter of the nuclear
region would have to be less than 1 kiloparsec. This nuclear region would be
about 100 times brighter than the luminous galaxies which have been identified
with radio sources so far...''.

The extreme luminosity of these sources, coming from very compact regions,
called for a new energy source. \citet{HoyleFowler1963b} were the first to
argue that such energy was of gravitational origin. \citet{Salpeter1964} and
\citet{Zeldovich1964} showed that accretion into a SMBH can indeed account for
the quasar luminosity. The idea was further elaborated by several authors
\citep[e.g.,][]{LyndenBell1969, LyndenBell1978, LyndenBellRees1971} and gained
widespread acceptance.

The argument went as follows. The total energy output from a quasar is at least
the energy stored in its radio halo ($\sim 10^{54}\,\hbox{J}=10^{61}\,$erg);
via $E=mc^2$ this corresponds to $10^7\,M_\odot$. Nuclear reactions have at
best an efficiency of 0.7\% (H burning). So the mass undergoing nuclear
reactions capable of powering a quasar is $> 10^9\,M_\odot$. Rapid variability
implies that a typical quasar is no bigger than a few light-hours. But the
gravitational energy of $10^9\,M_\odot$ compressed within this size is
$10^{55}\,$J, i.e. 10 times larger than the fusion energy. In Lynden-Bell's
words: ``Evidently, although our aim was to produce a model based on nuclear
fuel, we have ended up with a model which has produced more than enough energy
by gravitational contraction. The nuclear fuel has ended as an irrelevance.''

\citet{Salpeter1964} considered the radiative efficiency of accretion onto a
``Schwarzs\-child singularity'', showing that it can release an energy of
$0.057\,c^2$ per unit mass. \citet{Bardeen1970} showed that for a rotating
(Kerr) singularity up to nearly 42\% of the accreted rest mass energy is
emitted.


\section{BH demography}\label{sect:BHdemography}

\subsection{Why did it take so long to realize the AGN role in galaxy evolution?}

Although it was clear from the beginning that quasars are located in the nuclei
of galaxies, their presence was considered for decades just as an incidental
diversion, an ornament irrelevant for galaxy formation and evolution. There are
reasons for that:
\begin{itemize}

\item The physical scale of the AGNs is incomparably smaller than that of
    galaxies: typical radii of the stellar distribution of galaxies are of
    several kpc, to be compared with the Schwarzschild radius
\begin{equation}
    r_S = \frac{2 G\,M_{\rm BH}}{c^2} \simeq 9.56\times 10^{-6}\,\frac{M_{\rm
    BH}}{10^8\,M_\odot}\ \hbox{pc},
\end{equation}
     i.e. $r_{\rm BH} \sim 10^{-9}\,r_{\rm gal}$.

\item The radius of the "sphere of influence" of the SMBH (the distance at
    which its potential significantly affects the motion of the stars or of
    the interstellar medium) is also small:
\begin{equation}
    r_{\rm inf} = \frac{G\,M_{\rm BH}}{\sigma_\star^2} \simeq 11 \frac{M_{\rm BH}}{10^8\,M_\odot}\,
    \left(\frac{\sigma_\star}{200\,\hbox{km}\,\hbox{s}^{-1}}\right)^{-2}\ \hbox{pc},
\end{equation}
    $\sigma_\star$ being the velocity dispersion of stars in the host
    galaxies (SMBHs are generally associated to spheroidal components of
    galaxies, whose stellar dynamics is dominated by random motions, not by
    rotation). Hence SMBHs have a negligible impact on the global stellar and
    interstellar medium (ISM) dynamics.

\item AGNs and galaxies have very different evolutionary properties. AGNs
    evolve much faster: they are much rarer than galaxies at low redshifts,
    where galaxies were most extensively studied, and become much more
    numerous at $z \ge 2$.

\end{itemize}
Although powerful AGNs are rare locally, the SMBHs powering the quasars at high
$z$ do not disappear. After they stop accreting, they should live essentially
forever as dark remnants. So dead quasar engines should hide in many nearby
galaxies. This was pointed out early on \citep{LyndenBell1969, Schmidt1978},
but a direct test of this idea had to wait for decades.

\subsection{How solid is the evidence of SMBHs in galactic nuclei?}

The difficulty to reveal inactive SMBHs in galactic nuclei stems from the fact
that the SMBH masses required to power the AGNs are a tiny fraction of the
stellar mass (let alone the total mass, including dark matter!) of
galaxies\footnote{Estimates of $M_{\rm BH}/M_\star$ range from $\simeq 0.1\%$
\citep{Sani2011} to $\simeq 0.49\%$ \citep{KormendyHo2013}.}, and therefore
their radius of influence is very small. Even in nearby galaxies the angular
scale associated to $r_{\rm inf}$ is sub-arcsecond:
\begin{equation}
     \theta_{\rm inf} \sim 0.2\frac{M_{\rm BH}}{10^8\,M
     _\odot} \left(\frac{\sigma_\star}{200\,\hbox{km}\,\hbox{s}}\right)^{-2}
     \left(\frac{D}{10\,\hbox{Mpc}}\right)^{-1}\ \hbox{arcsec},
\end{equation}
where $D$ is the distance. Thus dynamical evidence for SMBHs is hard to find.

The first stellar dynamical SMBH detections followed in the mid- to late-1980s,
when CCDs became available on spectrographs and required the excellent seeing
of observatories like Palomar and Mauna Kea \citep[for reviews
see][]{KormendyRichstone1995, FerrareseFord2005}. The Hubble Space Telescope
(HST), by delivering five-times-better resolution than ground-based optical
spectroscopy, made it possible to find SMBHs in many more galaxies. This led to
the convincing conclusion that \textit{SMBHs are present in essentially every
galaxy that has a bulge component}.

Because of the smallness of their sphere of influence, resolving it is possible
only for relatively nearby, very massive SMBHs. Dynamical estimates based on
line widths may not be reliable because of contributions to the mass from other
components (dense star clusters, dark matter, ...).

Completely reliable estimates would require resolved proper motions of stars
surrounding the SMBH, but so far this could be achieved only for the Milky Way,
for which orbits have been determined for about 30 stars within $\simlt
0.5\,$arcsec from the SMBH and distances $\simlt 0.1\,$arcsec at periastron.
The SMBH mass can be derived for each of them. However a complete orbit is well
measured only for the star S2 which then provides the most accurate
determination of the SMBH mass: $M_{\rm BH}=4.30\pm 0.20\, ({\rm
statistical})\pm 0.30\,({\rm systematic})\times 10^6\,M_\odot$
\citep{KormendyHo2013}.

The pericenter radius of S2 is 0.0146\,arcsec. At the distance
$R_0=8.28\pm0.15\,({\rm statistical}) \pm 0.29\,({\rm systematic})\,{\rm kpc}$
\citep{Genzel2010}, this angular radius corresponds to 0.00059\,pc or $1,400\,
r_S$. The mass density within this radius is $\simeq 5\times
10^{15}\,M_\odot\,\hbox{pc}^{-3}$. The extended mass component within the orbit
of S2 (visible stars, stellar remnants and possible diffuse dark matter)
contributes less than 4 to 6.6\% of this central mass
\citep[$2\,\sigma$;][]{Gillessen2009}.

\citet{Maoz1998} investigated the dynamical constraints on alternatives to the
SMBH interpretation. Can the dark mass density be accounted for not by a point
mass but by an ultra-dense cluster of any plausible form of nonluminous
objects, such as brown dwarfs or stellar remnants? To answer this question he
has investigated the maximum possible lifetime of such dark cluster against the
processes of evaporation and physical collisions.  It is highly improbable that
clusters with a lifetime much shorter than the age of a galaxy ($\sim
10^{10}\,$yr, i.e. 10\,Gyr) survives till the present epoch. Thus short cluster
lifetimes lend strong support to the SMBH interpretation.

The mass and the lower limit to the mass density of the Milky Way central
object imply a cluster lifetime of only $\hbox{a\,few}\times 10^5\,$yr,
implying that the case for a SMBH is quite strong. The Milky Way situation is
however still unique. \citet{Maoz1998} found a cluster lifetime shorter than
$\simeq 10\,$Gyr for only another galaxy, NGC\,4258, for which VLBI
measurements at mas resolution (i.e. about 100 times better than the HST) of
the $H_2 O$ mega-maser at 22\,GHz from the circum-BH molecular gas disk were
available \citep{Miyoshi1995}.

The observations of \citet{Miyoshi1995} established the $H_2 O$ mega-masers as
one of the most powerful tools for measuring SMBH masses. At the distance of
NGC\,4258 1 mas corresponds to 0.035\,pc. The masers have almost exactly a
Keplerian velocity ($V\propto r^{-1/2}$), as expected in the case of a point
source gravitational field. So it is likely that the disk mass is negligible so
that the SMBH mass can be straightforwardly derived as $M_{\rm BH}= V^2 r/G$.

Useful maser disks are however rare, not least because they must be edge-on,
and the orientation of the host galaxy gives no clue about when this is the
case.  Also in several cases the disk mass was found to be comparable to the
SMBH mass, a major complication for SMBH mass measurements. However, progress
on this subject has been accelerating in recent years \citep{Kuo2017a,
Kuo2017b, Gao2017, Henkel2018}.

Information on the immediate environment, within a few gravitational radii, of
the SMBH is provided by the asymmetric K$\alpha$ iron line due to fluorescence
\citep[e.g.,][]{Fabian2013}. X-ray observations have shown that AGNs possess a
strong, broad Fe K$\alpha$ line at 6.4\,keV (rest-frame). Spectral-timing
studies, such as reverberation and broad iron line fitting, of these sources
yield coronal sizes, often showing them to be small and in the range of 3 to 10
gravitational radii in size \citep{Fabian2015}.  The line then experiences
general-relativistic effects, such as strong light bending and large
gravitational redshift. If the SMBH is rotating, its angular momentum manifests
through the Lense-Thirring precession which occurs only in the innermost part
of the accretion disk where the space-time becomes twisted in the same
direction that the SMBH is rotating.

The BH spin is measured by the parameter $a=c\,J/G\,M_{\rm BH}^2$, $J$ being
the angular momentum. Negative spin values represent retrograde configurations
in which the BH spins in the opposite direction to the disk, positive values
denote prograde spin configurations, and $a = 0$ implies a non-spinning BH. The
relativistic effects allow the BH spin to be measured. In particular, the Fe
K$\alpha$ line shape changes,  as a function of BH spin; the breadth of the red
wing of the line is enhanced as the BH spin increases \citep{Brenneman2013}.

\section{The So{\l}tan argument}\label{sect:Soltan}

A still not completely settled issue is: which is the main mechanism of SMBH
growth? Radiative accretion certainly contributes. Actually, most of the
accreted material is incorporated by the BH: the standard radiative efficiency
adopted by AGN evolutionary models is $\simeq 10\%$, implying that $\sim 90\%$
of the mass adds to the BH. On the other hand, models of merger-driven galaxy
evolution envisage that also the central SMBHs grow by merging. In this case
the angular momentum is dissipated by gravitational radiation, i.e. happens
without electromagnetic emission.

\citet{Soltan1982} pointed out that, if the SMBH growth is mostly due to
radiative accretion, the luminosity function of QSOs as a function of redshift
traces the accretion history of the SMBHs: for an assumed mass-to-energy
conversion efficiency, the luminosity function at any given redshift directly
translates into an accreted mass density at that redshift. Integrating such
mass density over redshift, gives a present day accreted mass density, which is
a lower limit to the present day SMBH mass density since the SMBH mass can have
also increased non-radiatively (e.g. via mergers).

Thus a comparison of the accreted mass density with the local SMBH mass density
provides considerable insight into the formation and growth of massive SMBHs.
If a BH is accreting at a rate \Mdot,\, its emitted luminosity is
\begin{equation}
L=\eff \Mdot_\mathrm{acc} c^2
\end{equation}
where $\eff$ is the radiation efficiency, i.e. the fraction of the accreted
mass which is converted into radiation and thus escapes the BH.

The growth rate of the BH, \Mdot, is thus given by
\begin{equation}\Mdot =(1-\eff) \Mdot_\mathrm{acc}.
\end{equation}
Let's neglect any process which, at time $t$, might `create' or `destroy' a BH
with mass $M$. In particular, \textit{this means neglecting BH merging}.
Indeed, the merging process of two BHs, $M_1+M_2\rightarrow M_{12}$, means that
BHs with $M_1$ and $M_2$ are destroyed while a BH with $M_{12}$ is created.
Then, if $\eff$ is constant we have:
\begin{equation}
\rhoBH = \frac{1-\eff}{\eff c^2} \, U_{T}
\end{equation}
where $U_{T}$ is the total {\it comoving} energy density from AGNs (not to be
confused with the total {\it observed} energy density), given by
\begin{equation}
U_T = \int_0^{{\zo}} dz \frac{dt}{dz} \int_{L_1}^{L_2} L \phi(L,z)\,d L\,.
\end{equation}
Here $\phi(L,z)\,d L$ is the \textit{bolometric} AGN luminosity function. Note
the factor $(1-\eff)$ which is needed to account for the part of the accreting
matter which is radiated away during the accretion process. If BH mergers,
which don't yield electromagnetic radiation but only gravitational waves, are
important, the derived  $\rhoBH$ is a lower limit.

\citet{Marconi2004} used luminosity functions in the optical B-band, soft X-ray
(0.5--2 keV) and hard X-ray (2--10 keV)  band \citep{Ueda2003} transformed into
a bolometric luminosity function using bolometric corrections obtained from
template spectral energy distributions (SEDs). In addition, they constrained
the redshift-dependent X-ray luminosity function to reproduce the hard X-ray
background, which can be considered an integral constraint on the total mass
accreted over the cosmic time and locked in SMBHs. They obtained:
\begin{equation}
\rhoBH=(4.7-10.6) \frac{(1-\eff)}{9\eff}\rhoBHunits,
\end{equation}
consistent with their own estimate from the local SMBH mass function,
$\rhoBH=(3.2-6.5)\rhoBHunits$, for the `canonical' value $\eff=0.1$.

A similar conclusion was reached by Ueda et al. (2014) using updated X-ray
luminosity functions, still constrained to reproduce the X-ray background, and
comparing with the local SMBH mass density by \citet{Vika2009}, $\rhoBH=(4.9
\pm 0.7) \rhoBHunits$,  derived from the empirical relation between SMBH mass
and host-spheroid luminosity (or mass).

However, recent analyses of SMBH mass measurements and scaling relations
concluded that the BH-to-bulge mass ratio shows a mass dependence and varies
from 0.1--0.2\% at $M_{\rm bulge} \simeq 10^9\,M_\odot$ to $\simeq 0.5\%$ at
$M_{\rm bulge} \simeq 10^{11}\,M_\odot$ \citep{GrahamScott2013,
KormendyHo2013}. A similarly large median $M_{\rm BH}/M_{\rm bulge}$ ratio for
\textit{early type galaxies} was found by \citet{Savorgnan2016} who however
reported a substantial decrease of the ratio with decreasing stellar mass of
the bulges of late-type galaxies.

The revised normalization is a factor of 2 to 5 larger than previous estimates
ranging from $\simeq 0.10\%$ \citep{MerrittFerrarese2001, McLureDunlop2002,
Sani2011} to $\simeq 0.23\%$ \citep{MarconiHunt2003}, therefore resulting in an
overall increase in the effective ratio, which is dominated by massive bulges.

This would imply either a lower mean radiative efficiency or an important
non-radiative (e.g. merging) contribution to the BH growth. Radiatively
inefficient processes (i.e. slim accretion disks) have been independently
advocated to explain the fast growth of SMBHs in the early Universe
\citep[e.g.,][]{Madau2014}.

On the other hand, it was pointed out that the bulge-disk decomposition can
lead to a considerable underestimate of the spheroid luminosity and stellar
mass, $M_\star$ \citep{SavorgnanGraham2016} and the selection bias (the AGNs
with higher luminosity, i.e., generally, with higher SMBH mass for given
stellar mass, are more easily detected) can lead to an overestimate of $M_{\rm
BH}$. Both effects lead to an overestimate of the $M_{\rm BH}/M_\star$ ratio.
According to \citet{Shankar2016} the selection bias leads to an overestimate of
$M_{\rm BH}$ by at least a factor of 3.

Based on a comprehensive analysis of the co-evolution of galaxies and SMBHs
throughout the history of the universe by a statistical approach using  the
continuity equation and the abundance matching technique, \citet{Aversa2015}
found a  mean $M_{\rm BH}$--$M_\star$ relation systematically lower by a factor
of $\simeq 2.5$ than that proposed by \citet{KormendyHo2013}. The
SMBH-to-stellar mass ratio was found to evolve mildly at least up to $z\simlt
3$, indicating that the SMBH and stellar mass growth occurs in parallel by in
situ accretion and star formation processes, with dry mergers playing a
marginal role at least for the stellar and SMBH mass ranges for which the
observations are more secure.

Although the issue is still debated, there are physical arguments indicating
that the AGN radiative efficiency, $\epsilon$, varies with the system age,
although how this happens is not yet clear. The global consistency between the
SMBH mass density inferred from the So{\l}tan approach and from the local SMBH
mass function constrains predictions on the gravitational wave signals expected
from SMBH mergers.

In the case of thin-disk accretion, $\epsilon$ may range from 0.057 for a
non-rotating BH to 0.32 for a rotating Kerr BH with spin parameter $a=0.998$
\citep{Thorne1974}. During a coherent disk accretion, the SMBH is expected to
spin up very rapidly, and correspondingly the efficiency is expected to
increase up to $\simeq 0.3$. On the other hand, when the mass is flowing
towards the SMBH at high rates (super-Eddington accretion), the matter
accumulates in the vicinity of the BH and the accretion may happen via the
radiatively-inefficient `slim-disk' solution \citep{Abramowicz1988,
Begelman2012, Madau2014, AbramowiczStraub2014, Volonteri2015} that speeds up
the SMBH growth.

This would relieve the challenge set by the existence of billion-solar-mass
black holes at the end of the reionization epoch \citep[see, e.g.,][for a
review]{Haiman2013}.  The most distant quasar discovered to date, ULAS
J$1120+0641$ at a redshift $z = 7.084$, is believed to host a black hole with a
mass of $2.0(+1.5, -0.7)\times 10^9\,M_\odot$, only 0.78\,Gyr after the big
bang \citep{Mortlock2011}. The need of a radiatively inefficient phase is
however debated. For example, \citet{Trakhtenbrot2017} argue that ``the
available luminosities and masses for the highest-redshift quasars can be
explained self-consistently within the thin, radiatively efficient accretion
disk paradigm''.

\citet{Trakhtenbrot2018} have observed with ALMA six luminous quasars at $z\sim
4.8$ finding spectroscopically confirmed companion sub-millimeter galaxies for
three of them.  The companions are separated by $\sim 14-45\,$kpc from the
quasar, supporting the idea that major mergers may be important drivers for
rapid, early BH growth. However, the fact that not all quasar hosts with
intense star formation are accompanied by interacting sub-mm galaxies, and
their ordered gas kinematics observed by ALMA, suggest that other processes may
be fueling these systems. They then conclude that data demonstrate the
diversity of host galaxy properties and gas accretion mechanisms associated
with early and rapid SMBH growth.

\section{Relationships between galaxies and central SMBHs}\label{sect:interaction}

A tight correlation between $M_{\rm BH}$ and the galaxy velocity dispersion,
$\sigma$, was reported, independently, by \citet{FerrareseMerritt2000} and by
\citet{Gebhardt2000}. Both papers claimed that the scatter was only 0.30 dex
over almost 3 orders of magnitude in $M_{\rm BH}$ and no larger than expected
on the basis of measurement errors alone. This suggested that the most
fundamental relationship between SMBHs and host galaxies had been found and
that it implies a close link between SMBH growth and bulge formation.

Many papers have expanded on this result with bigger samples \citep[for a
review, see][]{KormendyHo2013}. The $M_{\rm BH}$--$\sigma$ correlation has been
confirmed to be the strongest, with the lowest measured and intrinsic scatter
\citep{Saglia2016}. The correlations with the bulge mass, $M_{\rm bulge}$, is
also strong, except for the pseudo-bulge\footnote{While bulge properties  are
indistinguishable from those of elliptical galaxies, except that they are
embedded in disks, pseudo-bulges have more disk-like properties and are though
to be made by slow (``secular'') evolution internal to isolated galaxy disks.}
subsample .

\citet{Saglia2016} find, for ellipticals and classical bulges:
\begin{eqnarray}
\log(M_{\rm BH}/M_\odot)&=&(4.868\pm 0.32)\log(\sigma/\hbox{km}\,\hbox{s}^{-1})-(2.827\pm 0.75) \\
\log(M_{\rm BH}/M_\odot)&=&(0.846\pm 0.064)\log(M_{\rm bulge}/M_\odot)-(0.713 \pm 0.697)
\end{eqnarray}
Both relationships agree, within the errors, with those derived by
\citet{KormendyHo2013};  however, while \citet{Saglia2016} find  a mean $M_{\rm
BH}/M_{\rm bulge}$ ratio slowly decreasing with increasing $M_{\rm bulge}$,
\citet{KormendyHo2013} find a slowly increasing trend.

The normalization of the $M_{\rm BH}$--$M_{\rm bulge}$ relation is currently
debated. As mentioned above, the main issues are the difficulties of removing
the disk component to derive $M_{\rm bulge}$ and, even more, the bias on
$M_{\rm BH}$ estimates.

An inspection of the \citet{Saglia2016} results shows that:
\begin{itemize}
\item ``core'' ellipticals\footnote{Detailed studies of the central regions
    of early-type galaxies \citep{Ferrarese1994, Faber1997} have identified
    two distinct classes of galaxy centers: ``power-law'' galaxies, whose
    central surface brightness shows a steep power-law profile; and ``core''
    galaxies, where the luminosity profile turns over at a fairly sharp
    ``break radius'' into a shallower power-law. \citet{Ferrarese1994} and
    \citet{Faber1997} found evidence that global parameters of early-type
    galaxies correlated with their nuclear profiles: ``core'' galaxies tend
    to have high luminosities, boxy isophotes, and pressure-supported
    kinematics, while ``power-law'' galaxies are typically lower-luminosity
    and often have disky isophotes and rotationally supported kinematics.}
    have more massive BHs than other classical bulges, at a given $\sigma$ or
    bulge mass;

\item the smallest intrinsic and measured scatters of the $M_{\rm
    BH}$--$\sigma$ and $M_{\rm BH}$--$M_{\rm bulge}$ relations are measured
    for the sample of ``core'' ellipticals;

\item ``power-law'' early-type galaxies and classical bulges follow similar
    $M_{\rm BH}$--$\sigma$ and $M_{\rm BH}$--$M_{\rm bulge}$ relations;

\item pseudo-bulges have smaller SMBH masses than the rest of the sample at a
    given $\sigma$ or $M_{\rm bulge}$;

\item disks do not correlate with $M_{\rm BH}$ and it is unclear whether
    pseudo-bulges do: SMBH masses do not ``know about'' galaxy disks
    \citep{Kormendy2011}.

\end{itemize}
These relations suggest a tight link between star-formation activity in the
spheroidal components of galaxies (not in the disks) and SMBH growth. Such link
is strongly confirmed by the striking similarity of the evolution of the SMBH
accretion rate and of the star formation rate or of the AGN and galaxy
luminosity densities, especially at substantial redshifts where the star
formation mostly occurs in the spheroidal components \citep{Shankar2009,
Fiore2017}.

The growth histories of the stellar mass and of the AGN luminosity (hence of
SMBH mass)  share further similarities. The most massive early-type galaxies
form in short (duration $\simlt 1\,$Gyr), intense starbursts at high redshift
while less massive galaxies have more extended star formation histories that
peak later with decreasing mass \citep[e.g., Fig.\,9 of][]{Thomas2010}. This
``anti-hierarchical'' nature (called ``downsizing'') is mirrored in the SMBH
growth: the most massive SMBHs likely grow in intense quasar phases which peak
in the early universe, while less massive SMBHs have more extended, less
intense growth histories that peak at lower redshift \citep[e.g.,][]{Ueda2014}.
All that suggests a strong galaxy-AGN co-evolution.

\section{AGN impact on galaxy evolution}\label{sect:coev}

\subsection{AGN-driven winds}
As we have seen, the smallness of the radius of influence means that the SMBH's
gravity has a completely negligible effect on its host galaxy. On the other
hand, the energy released by the AGN
\begin{equation}
E_{\rm BH} \simeq \epsilon Mc^2 \sim 2\times 10^{61}{\epsilon \over 0.1}{M_{\rm BH}\over 10^8\, M_\odot}\,{\rm erg},
\end{equation}
where $\epsilon$ is the mass to radiation conversion efficiency, is  far larger
than the gas binding energy. Setting $M_{\rm gas} = f M_{\rm bulge}$, with
$f<1$, we have
\begin{equation}
E_{\rm gas} \sim {3\over 2} f M_{\rm bulge}\sigma^2 \sim 1.2 \times 10^{58}f {M_{\rm bulge}/M_{\rm BH}\over 10^3}{M_{\rm BH}\over 10^8\, M_\odot}\left({\sigma\over 200\,\hbox{km/s}}\right)^{-2}\,{\rm erg},
\end{equation}
where $\sigma$ is the line-of-sight velocity dispersion (the corresponding 3D
velocity is $v=\sqrt{3}\sigma$). This means that only a few percent of the SMBH
energy output may have a strong influence on the gas in the host galaxy,
potentially expelling it and, at the same time, limiting the SMBH own growth.

The SMBH energy release can potentially affect its surroundings in two main
ways. By far the stronger one (in principle) is through direct radiation. This
is particularly effective during heavily dust-obscured phases of AGN evolution
\citep{Fabian2002}.

The second form of coupling the SMBH energy release to a host bulge is
mechanical. The huge accretion luminosity of SMBH's may drive powerful gas
flows into the host, impacting into its interstellar medium. A well know form
of flow is jets, highly collimated flows driven from the immediate vicinity of
the SMBH (``radio-mode'' feedback). However to affect most of the bulge
requires a way of making the interaction relatively isotropic, perhaps with
changes of the jet direction over time. Moreover, radio observations show that
the jet energy is dissipated on scales from several kpc to Mpc, i.e. on scales
larger than that of a galaxy.  Therefore they are more relevant for heating the
intergalactic medium (IGM) in galaxy clusters.

A form of mechanical interaction that has automatically the right property are
near-isotropic winds carrying large momentum fluxes (``quasar mode'' feedback).
Such winds are indeed observed in many AGNs, as we will see.

\citet{SilkRees1998} pointed out that an AGN emitting at close to the Eddington
rate could expel gas completely from its host galaxy provided that
\begin{equation}M_{\rm BH} > {{f\sigma^5\sigma_{\rm T}}\over{4\pi G^2 m_{\rm p} c}},
\end{equation}
where $\sigma_{\rm T}$, $G$, $m_p$, $c$ and $f$ are the Thomson cross section
for electron scattering, the gravitational constant, the proton mass, the speed
of light and the gas mass fraction, respectively. The galaxy bulge is assumed
to be isothermal with the radius $r$, so that its mass is $M_{\rm
bulge}=2\sigma^2 r/G$. The gas mass can be written as $M_{\rm gas} = f M_{\rm
bulge}$ with $f<1$. The maximum collapse rate of the gas is $\dot{M}_{\rm gas}=
{M_{\rm gas}/t_{\rm free-fall}}$ with $t_{\rm free-fall}=r/\sigma$. The
corresponding power is
\begin{equation}\dot{E}_{\rm gas}= {1/2}({M_{\rm gas}/t_{\rm free-fall}})v^2 = 3{f\sigma^5/G},
\end{equation}
($v=\sqrt{3}\sigma$). The relation $M_{\rm BH}$--$\sigma$ then follows equating
$\dot{E}_{\rm gas}$ to the Eddington luminosity
\begin{equation}L_{\rm Edd}={4\pi G M_{\rm BH}m_p c\over \sigma_T}.
\end{equation}
Plugging in the numbers and considering that only a fraction, $f_{\rm Edd}$, of
the Eddington luminosity can be used to throw the gas out we get:
\begin{equation}M_{\rm BH}={3.6\times 10^5 \over f_{\rm Edd}}\left({\sigma\over 100\,\hbox{km/s}}\right)^5
\,M_\odot.
\end{equation}
A comparison with the empirical $M_{\rm BH}$--$\sigma$ relation shows that
$f_{\rm Edd}$ at the few/several percent level is enough to stop the accretion
and to expel the gas from the host galaxy.

Alternatively, we may have momentum or force (instead of energy) balance
\citep{Fabian1999, Fabian2002, King2003, King2005, Murray2005}. Balancing the
outward radiation force with the inward one due to gravity gives
\begin{equation}{{4\pi GM_{\rm BH}m_{\rm p}}\over \sigma_{\rm T}}=
{L_{\rm Edd}\over c}={{GM_{\rm gal}M_{\rm gas}}\over
  r^2}={{fGM_{\rm gal}^2}\over r^2}={{fG}\over r^2}{\left({2\sigma^2
    r}\over G\right)^2}\end{equation}
i.e.,
\begin{equation}{{4\pi G M_{\rm BH}m_{\rm p} }\over{\sigma_{\rm T}
  }}={{4f\sigma^4}\over G},
\end{equation}
from which we get
\begin{equation}
M_{\rm BH}={{f\sigma^4\sigma_{\rm T}}\over{\pi G^2 m_{\rm p}}}=
{1.43\times 10^9}f \left({\sigma\over 100\,\hbox{km/s}}\right)^4\,M_\odot.
\end{equation}
Thus in the case of momentum-driven flows, the SMBH mass required to stop the
accretion is a factor $\sim c/\sigma$ larger than in the energy-driven case. A
comparison with the empirical $M_{\rm BH}$--$\sigma$ relation shows that in
this case, even the full Eddington luminosity is not enough to unbind the gas
unless the gas fraction is small ($f < 0.1$) or the AGN has a strongly
super-Eddington luminosity. That the full AGN luminosity goes into radiation
pressure is expected in the case of ``Compton-thick'' objects, most of whose
radiation is absorbed by the gas.

In the radiation-driven case it is implicitly assumed that the cooling of the
SMBH wind is negligible.  Under what conditions does this apply? We can crudely
model the outflows as quasi-spherical winds from SMBHs accreting at about the
Eddington rate
\begin{equation}
\dot{M}_w\simeq \dot{M}_{\rm Edd}= {L_{\rm Edd}\over \epsilon c^2}\simeq 0.22
{M_{\rm BH}\over 10^8}\,M_\odot\,\hbox{yr}^{-1}.
\end{equation}
Winds like this have electron scattering optical depth $\tau\sim 1$, measured
inward from infinity to a distance of order the Schwarzschild radius $R_{\rm
S}=2 G M/ c^2$ \citep{KingPounds2015}. So on average every photon emitted by
the AGN scatters about once before escaping to infinity.

Because electron scattering is front-back symmetric, each photon on average
gives up all its momentum to the wind, and so the total (scalar) wind momentum
should be of order the photon momentum, or
\begin{equation}
\dot{M}_w v\sim {L_{\rm Edd}\over c}\simeq \dot{M}_{\rm Edd} \epsilon c,
\end{equation}
where $v$ is the wind’s terminal velocity. Since $\dot{M}_w\simeq \dot{M}_{\rm
Edd}$ we get \citep{KingPounds2015}:
\begin{equation}
v\simeq 0.1 {\epsilon\over 0.1} c\, .
\end{equation}
The instantaneous wind mechanical luminosity is then
\begin{equation}
L_{\rm BHwind}={1\over 2}v^2\dot{M}_w\simeq {1\over 2}{v\over c} L_{\rm Edd}
\simeq 0.05 {\epsilon\over 0.1} L_{\rm Edd},
\end{equation}
in good agreement with the earlier estimate for the energy-driven winds.

A self-consistent model is described by \citet{KingPounds2015}. The black hole
wind is abruptly slowed in an inner (within the SMBH sphere of influence)
shock, in which the temperature approaches $\sim 10^{11}\,$K. The shocked wind
gas acts like a piston, sweeping up the host ISM at a contact discontinuity
moving ahead of it. Because this swept-up gas moves supersonically into the
ambient ISM, it drives an outer (forward) shock into it. The dominant
interaction here is the reverse shock slowing the black hole wind, which
injects energy into the host ISM. The nature of this shock differs sharply
depending on whether some form of cooling (typically radiation) removes
significant energy from the hot shocked gas on a timescale shorter than its
flow time.

If the cooling is strong  (momentum-driven flow), most of the pre-shock kinetic
energy  is lost (usually to radiation). As momentum must be conserved, the
post-shock gas transmits just its ram pressure to the host ISM. As we have
seen, this amounts to transfer of only a fraction $\sim \sigma/c\sim  10^{-3}$
of the mechanical luminosity $L_{\rm BH,wind}\simeq 0.05\,L_{\rm Edd}$ to the
ISM. In other words, for SMBHs close to the $M_{\rm BH}$--$\sigma$ relation, in
the momentum-driven limit only $\sim 10\%$ of the gas binding energy  is
injected into the bulge ISM, which is therefore stable.

In the opposite limit in which cooling is negligible, the post-shock gas
retains all the mechanical luminosity and expands adiabatically into the ISM.
The post-shock gas is now geometrically extended. The mechanical energy,
thermalized in the shock and released to the ISM, now equals the gas binding
energy. The energy-driven flow is much more violent than the momentum-driven
flow and can unbind the bulge.

{Note that if the SMBH and galaxy evolve strictly in parallel, preserving the
$M_{\rm BH}$--$\sigma$ relation, a SMBH in an energy-driven environment is
unlikely to reach the observed SMBH masses because it would stop its gas
accretion}. We will come back to this point in the following.

To sum up, for both momentum-  and energy-driven outflows a fast wind (velocity
$\sim 0.1c$) impacts the interstellar gas of the host galaxy, producing an
inner reverse shock that slows the wind and an outer forward shock that
accelerates the swept-up gas. In the momentum-driven case, the shocks are very
narrow and rapidly cool to become effectively isothermal; only the ram pressure
is communicated to the outflow, leading to very low kinetic energy, $\sim
(\sigma/c)\,L_{\rm Edd}$. In an energy-driven outflow, the shocked regions are
much wider and do not cool; they expand adiabatically, transferring most of the
kinetic energy of the wind to the outflow.

\subsection{Observed wind properties}

A recent study of relations between AGN properties, host galaxy properties, and
AGN winds has been carried out by \citet{Fiore2017}. This paper also contains
an exhaustive list of references to observations of massive outflows of
ionised, neutral and molecular gas, extended on kpc scales, with velocities of
order of $1000\,\hbox{km}\,\hbox{s}^{-1}$. Three main techniques have been used
to detect such outflows \citep[see][for references]{Fiore2017}: deep
optical/near-infrared spectroscopy, mainly from integral field observations;
interferometric observations in the (sub)millimetre domain; far-infrared
spectroscopy from \textit{Herschel}. In addition, AGN-driven winds on
sub-parsec scales, from the accretion disk scale up to the dusty torus, are now
detected routinely up to $z>2$ as blue-shifted absorption lines in the X-ray
spectra of a substantial fraction of AGNs \citep[e.g..][]{Kaastra2014}. The
most powerful of these winds have extreme velocities (ultrafast outflows, UFOs,
with $v\sim 0.1$--$0.3c$) and are made by highly ionised gas which can be
detected only at X-ray energies.

The main conclusions of the \citet{Fiore2017} study are:
\begin{itemize}

\item the mass outflow rate is correlated with the AGN bolometric luminosity;

\item the fraction of outflowing gas in the ionised phase increases with the
    bolometric luminosity;

\item the wind kinetic energy rate (kinetic power) $\dot{E}_{\rm kin}$ is
    correlated with the AGN bolometric luminosity, $L_{\rm bol}$, for both
    molecular and ionized outflows: we have $\dot{E}_{\rm kin}/L_{\rm
    bol}\sim 1 - 10\%$ for molecular winds and $\dot{E}_{\rm kin}/L_{\rm
    bol}\sim 0.1 - 10\%$ for ionised winds;

\item About half X-ray absorbers and broad absorption line (BAL) winds have
    $\dot{E}_{\rm kin}/L_{\rm bol}\sim 0.1 - 1\%$ with another half having
    $\dot{E}_{\rm kin}/L_{\rm bol}\sim 1 - 10\%$.

\item Most molecular winds and the majority of ionised winds have kinetic
    power in excess to what would be predicted if they were driven by
    supernovae, based on the SFRs measured in the AGN host galaxies. The
    straightforward conclusion is that the most powerful winds are AGN
    driven.

\item The average AGN wind mass-loading factor\footnote{The mass-loading
    factor, $\eta$, is the ratio between the mass outflow rate, $\dot{M}_w$,
    and the SFR: $\eta=\dot{M}_w/\hbox{SFR}$.}, $\langle \eta \rangle$, is
    between 0.2 and 0.3 for the full galaxy population while $\langle \eta
    \rangle \gg 1$ for massive galaxies at $z \simlt 2$. We may then
    tentatively conclude that AGN winds are, on average, powerful enough to
    clean galaxies from their molecular gas (either expelling it from the
    galaxy or by destroying the molecules) in massive systems only, and at $z
    \simlt 2$.

\item What happens at $z> 2$ is still unclear.

\end{itemize}
AGN winds may then be the manifestation of AGN feedback linking nuclear and
galactic processes in massive galaxies and accounting for the correlation of
SMBH masses and properties of host bulges.

\subsection{Do SFR and BH accretion rate simply track each other?}

A widely cited galaxy-AGN co-evolution model \citep[][see, in particular, their
Fig.~1]{Hopkins2008} envisage the following scenario for galaxy evolution.
Early galaxies have a disk morphology and grow mainly in quiescence until the
onset of a major merger. During the early stages of the merger, tidal torques
excite some enhanced star formation and SMBH accretion. During the final
coalescence of the galaxies, massive inflows of gas trigger strong  starbursts.
The high gas densities feed a rapid SMBH growth.

In this scenario, star formation and SMBH accretion evolve strictly in
parallel. If so we expect a direct proportionality between the corresponding
luminosities, i.e., in practice between infrared (IR)\footnote{It has become
common practice to define the IR luminosity, $L_{\rm IR}$, as the integrated
emission between 8 and $1000\,\mu$m. In this paper we adopt this convention.}
and X-ray luminosities, $L_{\rm X}$. In fact, since star formation at
substantial redshifts is dust enshrouded, the light emitted by young stars is
absorbed by dust and re-emitted at far-IR wavelengths; so $L_{\rm IR}$ is a
measure of the luminosity produced by star formation. In turn, $L_{\rm X}$ is
the best indicator of accretion luminosity.

The connection between star formation and SMBH accretion has been investigated
by several authors using samples of far-IR selected galaxies followed up in
X-rays and of X-ray/optically selected AGNs followed up in the far-IR band
\citep[e.g.,][and references therein]{Lapi2014}. Most recently,
\citet{Lanzuisi2017} looked for correlations between the average properties of
X-ray detected AGNs and their far-IR detected star forming host galaxies, using
a large sample of X-ray and far-IR detected objects in the COSMOS field. The
sample covered the redshift range $0.1< z <4$ and about 4 orders of magnitude
in X-ray and far-IR luminosity, and in stellar mass. They found that $L_{\rm
X}$ and $L_{\rm IR}$ are significantly correlated. However, splitting the
sample into five redshift bins, they found that for every redshift bin both
luminosities have a broad distribution, with weak or no signs of a correlation
(see their Fig.\,4).

\citet{Lanzuisi2017} further investigated, for each redshift bin, the
relationships between $L_{\rm IR}$ and $L_{\rm X}$ in $L_{\rm X}$ bins and
between $L_{\rm X}$ and $L_{\rm IR}$ in $L_{\rm IR}$ bins (see their Fig.\,5).
The two relationships can be very different if the link between the two
luminosities is complex, as in the case of a weak correlation for the bulk of
the population and a strong correlation only for the most extreme objects. This
turns out to be case. The average host $L_{\rm IR}$ has a quite flat
distribution in bins of $L_{\rm X}$ (i.e. the two quantities are weakly, if at
all, correlated), while the average $L_{\rm X}$ somewhat increases in bins of
$L_{\rm IR}$ with logarithmic slope of $\simeq 0.7$ in the redshift range $0.4
< z <1.2$. At higher redshifts the slopes become flatter.

The simplest interpretation of these data goes as follows. At high-$z$ the gas
is very abundant in galactic halos and therefore both the star formation and
the SMBH accretion can proceed vigorously. However the timescales are widely
different. In the case of star formation, the relevant timescale is the minimum
between the dynamical time, which is $\sim 0.1\,$Gyr for massive galaxies, and
the cooling timescale, that can be much longer. Also, since there is an
abundant amount of gas available, the star formation can proceed until the gas
is swept outside the galaxy by some feedback process (see
Sect.~\ref{sect:feedback}). As argued by several authors \citep{Granato2004,
Thomas2010, Lapi2014} the star formation timescale is likely of at least
0.5--0.7\,Gyr for massive spheroidal galaxies and longer for less massive
galaxies.

The SMBH accretion timescale is much shorter. During the early evolutionary
phases the AGN accretion rate likely occurs at about the Eddington limit:
\begin{equation}
L_{\rm AGN}=\epsilon c^2 \dot{M}_{\rm BH}=\lambda L_{\rm Edd}
\end{equation}
where $\lambda$ is the Eddington ratio and
\begin{equation}
L_{\rm Edd}= \frac{4\pi c G M_{\rm BH} \mu_e}{\sigma_T} =
1.51\times 10^{38}\frac{M_{\rm BH}}{M_\odot}\ \hbox{erg}\,\hbox{s}^{-1}
\end{equation}
$\mu_e\simeq 1.2 m_p$ being the mass per unit electron and $\sigma_T$ the
Thomson cross-section.

If $L_{\rm AGN}= L_{\rm Edd}$, i.e. $\lambda=1$, $M_{\rm BH}$ grows
exponentially:
\begin{equation}
M_{\rm BH}=M_{\rm seed} e^{t/\tau_{\rm S}}
\end{equation}
where $\tau_{\rm S}$ is the Salpeter time
\begin{equation}
\tau_{\rm S}=\frac{\epsilon c \sigma_T}{4\pi G \mu_e}= 3.7\times 10^7 \frac{\epsilon}{0.1}\, \hbox{yr}.
\end{equation}
Because of the large difference between the accretion and the star-formation
timescale, during \textit{the Eddington-limited regime} the star-formation and
the AGN luminosities cannot be proportional to each other: the AGN luminosity
increases exponentially while that due to star formation is approximately
constant.

On the other hand, at later times, the accretion is strongly sub-Eddington and
the accretion timescale can match the star-formation timescale, as in the case
of re-activation of star-formation and nuclear activity (``rejuvenation'') by
interactions or mergers.

\section{Feedback and galaxy evolution}\label{sect:feedback}

The AGN feedback avoids the formation of too many massive galaxies and the
presence of too many baryons in their galactic halos. In fact, while the mean
cosmic baryon density in units of the critical density is $\Omega_{\rm b}
=0.0486$ \citep{PlanckCollaborationXIII2016}, the baryon density in galaxies is
$\Omega_{\rm b, gal} \simeq 0.00345$ \citep{FukugitaPeebles2004}, i.e. only
$\sim 7\%$ of baryons are in galaxies. Thus a process capable of sweeping out
more that 90\% of baryons initially present in galactic halos is necessary.

There are however other structural problems that cannot be solved by the AGN
feedback alone. The na\"{\i}ve assumption that the stellar mass follows the
halo mass leads also to too many small galaxies \citep[see Fig. 1
of][]{SilkMamon2012}. For halo masses $\simlt 10^{11}\,M_\odot$ the energy
input into the interstellar medium is dominated by supernovae
\citep{Granato2004} that become the key player in determining the slope of the
low-mass portion of the galaxy stellar mass function as well as the shape of
the Faber-Jackson and of the $M_{\rm BH}$--$\sigma$ for less massive bulges.

Why is the energy injection by supernovae insufficient for the most massive
galaxies? Although the \textit{total energy} released by supernovae, integrated
over the galaxy lifetime, may be large, the mean \textit{power} (energy
released per unit time) is not enough to expel the gas from large galaxies.
According to \citet{DekelSilk1986} supernovae can cause gas disruption and
dispersal in intermediate mass and massive dwarf galaxies (halo mass $\sim 10^8
- 10^{10}\,M_\odot$) and can expel the remaining baryons in systems of halo
mass up to $\sim 10^8\,M_\odot$, leaving behind dim dwarf galaxy remnants.
According to \citet{Granato2004}, if 5\% of the energy released by supernova
explosions goes into heating of the gas, supernova-driven winds can eject large
gas fractions from halos of up to $\sim 10^{11}\,M_\odot$.

In very low-mass halos gas cannot even fall in, because its specific entropy is
too high \citep{Rees1986}. Only halos of mass $> 10^5\,M_\odot$ can trap
baryons that are able to undergo early $\hbox{H}_2$ cooling and eventually form
stars. Heating by re-ionization increases this mass limit. The abrupt increase
of the sound speed to $10 - 20\,\hbox{km}\,\hbox{s}^{-1}$ at $z \sim 8$ means
that dwarfs of halo mass $\sim 10^6 - 10^7\,M_\odot$, which have not yet
collapsed and fragmented into stars, will be disrupted \citep{SilkMamon2012}.


\section{Conclusions}\label{sect:conclusions}

SMBH masses are tightly correlated with properties of the spheroidal component
(classical bulges) of their host galaxies, such as the stellar velocity
dispersion (that shows the tightest correlation), the stellar mass, the
luminosity. In contrast, they do not correlate with disk properties.

Since the spheroidal components of galaxies possess the older stellar
populations, this suggests that the SMBH growth happens  in the context of
dissipative baryon collapse at substantial redshifts. The evolution of
luminosity densities due to star-formation in spheroidal galaxies and to AGN
activity proceed in parallel, i.e. the SMBH growth and galaxy build up follow a
similar evolution through cosmic history, with a peak at $z\sim 2 - 3$ and a
sharp decline toward the present age.

This implies a mutual relationship beween star formation  and SMBH accretion
rates, which however do not simply track each other. Most growth of large SMBHs
happens by radiatively efficient gas accretion (So{\l}tan argument). The energy
radiated by a SMBH ($\sim 0.1\,M_{\rm BH} c^2$, assuming a radiative efficiency
of 10\%) is much larger than the binding energy of its host bulge ($\sim M_{\rm
bulge} \sigma^2$, $\sigma$ being the stellar velocity dispersion). If only
$\sim 5\%$ of the AGN energy output couples to gas in the forming galaxy, then
all of the gas can be blown. Thus, SMBH growth may be self-limiting, and AGNs
may quench star formation.

The substantial stellar masses and star-formation rates of sub-millimeter
galaxies (SMGs) and the evidence for subdominant AGN activity and moderate SMBH
masses in these objects imply that most of the star formation occurs before the
SMBH reach large masses. This is easily understood since the early SMBH growth
is Eddington limited.

When the SMBHs reach a critical threshold, their ``quasar-mode energy
feedback'' balances outward radiation or mechanical pressure against gravity.
Then the AGNs blow away the interstellar gas, quenching star formation and
leaving the galaxies red and dead. AGNs become visible and continue to shine
for a few Salpeter times, accreting the ``reservoir'' (torus) mass. This
convincingly solves some problems of galaxy formation, such as expelling a
large fraction of initial gas to account for the present day baryon to dark
matter ratio in galaxies and preventing excessive star formation (the stellar
mass function of galaxies sinks down,  at large masses, much faster than the
halo mass function).

However, observations of winds capable of removing the interstellar gas from
$z> 2$ galaxies are still missing, and a large fraction of massive early type
galaxies formed most of their stars at these redshifts. To remove the gas we
need a kinetic power of $\sim 5\%$ of the AGN bolometric luminosity, but the
scanty observations of high $z$ winds generally indicate lower kinetic powers.

Two kinds of AGN feedback have been considered. Radio-mode feedback (powerful
jets of radio sources) is a well known phenomenon but can hardly substantially
affect the galaxy evolution: it is very hard to confine well-collimated jets
within their galaxies. As \cite{KormendyHo2013} put it: ``Firing a rifle in a
room does not much heat the air in the room''. It is much more plausible that
the jet energy is dissipated by interactions with the intergalactic gas,
especially in galaxy clusters.

To efficiently operate on galaxy scales, the feedback must be relatively
isotropic (quasar-model feedback). But the physics is not well understood. The
correlations between SMBH masses and the galaxy velocity dispersions are
consistent with both energy- and momentum-driven feedback. But only
energy-driven feedback can efficiently quench star formation.

Outflows with kinetic power sufficient to clean the galaxy of cold gas were
found for only a few high-$z$ galaxies. They are more common at $z \simlt 2$.
But most local AGNs accrete at very sub-Eddington rates. Very few galaxies are
still growing their SMBHs at a significant level. Rapid SMBH growth by
radiatively efficient accretion took place mostly in more massive galaxies that
are largely quenched today. That is, the era of SMBH growth by radiatively
efficient accretion is now mostly over: co-evolution happened at high $z$.

\acknowledgements{I'm grateful to the the organizers of the Third Cosmology
School in Cracow for the kind invitation and the extraordinarily warm
hospitality.  Work supported in part by ASI/INAF agreement n.~2014-024-R.1 for
the {\it Planck} LFI Activity of Phase E2 and   }

\bibliographystyle{ptapap}
\bibliography{ptapapdoc}

\end{document}